\documentclass[12pt,a4paper,final]{iopart}
\usepackage[]{graphicx}
\usepackage{amsmath}
\usepackage{amsfonts}
\usepackage{color}
\usepackage{graphicx}
\usepackage{amsbsy}
\usepackage{epstopdf}
\usepackage{rotating}
\usepackage{cancel}

\def\ket#1{| #1 \rangle}

\def\Tr{{\mathrm{Tr}}}

\newcommand{\dg}{^\dagger}

\newcommand{\ketbra}[2]{ | #1 \rangle \langle #2 | }

\definecolor{gold}{rgb}{0.85,.66,0}
\definecolor{brown}{rgb}{0.647,0.165,0.165}

\date{\today}

\begin{document}

\title{Generating nonclassical states of motion using spontaneous emission}
\author{Ben Q. Baragiola}
\address{Centre for Quantum Computation and Communication Technology, School of Science, RMIT University, Melbourne, Victoria 3001, Australia}

\author{Jason Twamley}
\address{Centre for Engineered Quantum Systems, Department of Physics and Astronomy, Macquarie University, Sydney, NSW 2109, Australia}

\begin{abstract}
Nonclassical motional states of matter are of interest both from a fundamental perspective but also for their potential technological applications as resources in various quantum processing tasks such as quantum teleportation, sensing, communication, and computation. In this work we explore the motional effects of a harmonically trapped, excited two-level emitter coupled to a one-dimensional (1D) photonic system. %
As the emitter decays it experiences a momentum recoil that entangles its motion with the emitted photon pulse.  
In the long-time limit the emitter relaxes to its electronic ground state, while its reduced motional state remains entangled with the outgoing photon. 
We find photonic systems where the long-time reduced motional state of the emitter, though mixed, is highly nonclassical and in some cases approaches a pure motional Fock state.  
Motional recoil engineering can be simpler to experimentally implement than complex measurement and feedback based methods to engineer novel quantum mechanical states of motion.
\end{abstract}


\vspace{2pc}
\section{\label{sec:intro} Introduction}

Engineering nonclassical quantum states of motion has long attracted much attention particularly if the motional properties of the object in question admit the potential to generate macroscopic superpositions. Early work, both theoretical and experimental, on the creation of nonclassical motional states focused on the motional states of individual trapped ions in the late 90s \cite{Cirac1993b, Meekhof1996, Gou1996a, Gou1996h, Gou1996f, Bardroff1996, Steinbach1997d}.
Much of this work was performed in the Lamb-Dicke (LD) regime where the photon recoil energy is much smaller than a motional quanta, and thus the ion suffers little motional excitation. 
Beyond this regime (so called large LD regime), schemes for cooling via electromagnetically induced transparency \cite{Roghani2008} and the existence of bistable motional states \cite{Wang2008} have been proposed. Studies involving trapped ions are typically done in a free-space setting. 
Alternatively, strong coupling of quantum systems to 1D electromagnetic environments, such as optical waveguides, provides a new avenue to study light-matter interactions. 
Theoretical studies have shed light on the scattering of single-photon pulses \cite{Shen2005}, waveguide-mediated entanglement \cite{Zheng2013}, ultra-strong coupling \cite{Sanchez-Burillo2014}, and spatial light rectification \cite{Dreisow2013, Fratini2014, Dai2015, Mascarenhas2016}. However, little theoretical research has been done on the optomechanical effects of mobile emitters ina 1D photonic system, even while much experimental progress has been made towards strong coupling of optically trapped emitters \cite{Goban2014, Kato2015, Goban2015}. A notable exception is the recent work of Li \emph{et al.} \cite{Li2013}, 
which studied the modification of single-photon reflection spectra due to motional recoil of a trapped two-level emitter near a 1D waveguide. 
Complementary studies \cite{Jia2013, Ren2013b, Wang2015} use a standard phenomenological optomechanical coupling to include motional effects, including the creation of nonclassical motional states \cite{Milburn2016b, Hoff2016}. In the following we focus on microscopic optomechanical systems where an emitter, \emph{e.g.} an atom or nanoparticle containing a emitter, is held in a harmonic motional trap near to a 1D photonic system. The motional coupling is fundamentally mediated by the recoil suffered by the emitter upon emission/absorption of a photon, and no phenomenology need be assumed for the optomechanical interaction.  

In this article we explore the motional effects of spontaneous emission for an excited, harmonically trapped quantum emitter both within and beyond the LD regime. Optical decay into the modes of a bidirectional waveguide couples the emitter's motional state to the outgoing photon. In the long-time limit all the light escapes away along the waveguide, and the emitter relaxes to its electronic ground state. However, during the emission process the motional state is disturbed indefinitely. In the long-time limit entanglement between the motion and the outgoing photonic field produces reduced motional states that are often highly mixed and lack nonclassical features. We examine this pedagogical case in Sec. \ref{sec:waveguide}. We then explore two alternate physical architectures where the reduced motional state can become highly nonclassical. In Sec. \ref{Sec:mirror}  we consider an emitter next to a perfect mirror embedded in the waveguide. Including the motional coupling significantly modifies the well-known inhibition of decay due to interference from the field reflected off the mirror. We find that this effect can be used to generate pure single-excitation Fock states of motion. 
The second physical architecture involves coupling the harmonically trapped emitter to the circulating modes of a toroidal optical cavity, which is symmetrically coupled to left- and right-propagating modes in the waveguide. This architecture has been experimentally realized with tapered optical nanofibers by Rauschenbeutel and co-workers who have demonstrated a variety of effects including strong coupling \cite{Volz2014}, optical switching \cite{OShea2013}, and optical isolation \cite{Sayrin2015}. To date the effects of recoil on the trapped emitter have not been experimentally investigated. We explore this architecture in Sec. \ref{sec:toroidal}, and identify a parameter regime in which the reduced motional state is both highly entangled with the emitted photonic field and also individually nonclassical. 

For each physical scenario a quantum system couples to a bidirectional waveguide that supports a continuum of modes for each propagation direction.  These modes behave as an electromagnetic reservoir into which system excitations can decay. This open-systems description of light-matter interactions is conveniently treated within the context of input-output theory \cite{GardZoll04}. We consider the case where there are no initial optical fields present and there are no intrinsic losses, \emph{i.e.} the combined motional/optical system evolves unitarily.  Then, tracing over the two waveguide channels gives the dynamics of the reduced quantum system in terms of a Markov master equation,
\begin{align} \label{eq:masterequation}
	\frac{d}{dt} \hat{\rho} = - i [\hat{H}_{\rm sys}, \hat{\rho}] + \sum_{i=\ell,r} \hat{L}_i \hat{\rho} \hat{L}_i^\dagger - \frac{1}{2}  \hat{L}_i^\dagger \hat{L}_i \hat{\rho} - \frac{1}{2}  \hat{\rho} \hat{L}_i^\dagger \hat{L} _i,
\end{align}
where the subscripts $\{\ell,r\}$ refer to the left- and right-propagating waveguide modes, respectively.
The time evolution proceeds according to a system Hamiltonian $\hat{H}_{\rm sys}$ and two jump operators $\hat{L}_i$, one for each propagating mode. The jump operators describe the coupling of the quantum system to the left- and right-propagating modes.  For each of the physical scenarios below we solve the master equation numerically for the system state $\hat{\rho}(t)$ and then trace out additional system degrees of freedom to find the reduced state of the motion, $\hat{\rho}_m(t) = \Tr_{\mathcal{H}_0} [\hat{\rho}(t)]$, where $\mathcal{H}_0$ is the Hilbert space of all other system degrees of freedom.

\section{Trapped atom directly coupled to a waveguide} \label{sec:waveguide}

\begin{figure}
\centering
\includegraphics[width=13cm]{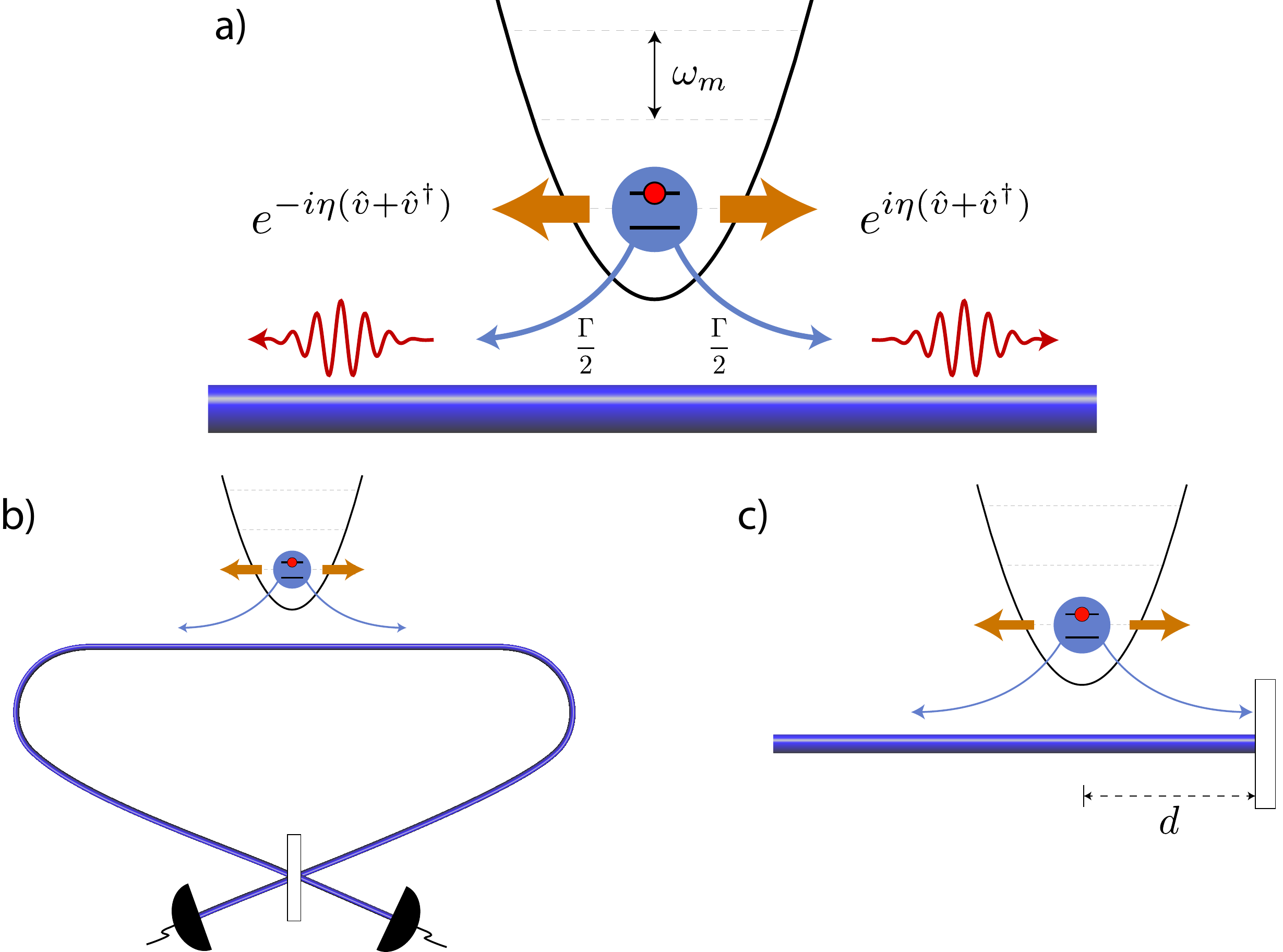}
\caption{Excited two-level quantum emitter decaying with total rate $\Gamma$ into a nearby optical waveguide. The emitter is in the motional ground state of a harmonic trap with frequency $\omega_m$ and phonon annihilation operator $\hat{v}$. Photon emission is accompanied by a momentum recoil with Lamb-Dicke parameter $\eta$. (a) The emitter decays symmetrically into right- and left-propagating optical modes of the waveguide.  The emitted optical fields travel off to $x\rightarrow \pm \infty$. (b) Conditional evolution: the emitted optical fields are redirected onto a 50/50 beamsplitter and detected. (b) A perfect mirror is placed a distance $d$ from the equilibrium position of the trap. }
\label{WaveguideSetup}
\end{figure}

We first consider a trapped quantum emitter strongly coupled to a bidirectional waveguide as shown in Fig.~\ref{WaveguideSetup}a. While this scenario reveals no remarkable nonclassical features in the unconditional long-time motional state of the emitter, it serves as an instructive launching point for the investigations below. Further, by considering the \emph{conditional} state of motion given detection of the emitted photon, as shown in Fig.~\ref{WaveguideSetup}b, we gain insight into physical settings that can indeed produce nonclassical signatures in the reduced motional state. Based on this we later study the case when the emitter is placed in front of a mirror as depicted in Fig.~\ref{WaveguideSetup}c.

The emitter is taken to be a two-level system with internal electronic eigenstates $\ket{g}$ and $\ket{e}$ separated by transition frequency $\omega_0$.  The emitter is held in a harmonic trap with a zero-point motion amplitude $x_{zpm}$, and bosonic operators $\hat{v}$, and $\hat{v}\dg$, that respectively annihilate and create motional quanta at frequency $\omega_m$ and satisfy $[\hat{v}, \hat{v}\dg] = 1$. We work in the interaction picture with respect to the electronic degree of freedom, so the Hamiltonian is that of harmonic motion in the trap, $\hat{H}_{\rm sys} = \omega_{m} \hat{v}^\dagger \hat{v}.$ The emitter decays symmetrically into the left- and right-propagating waveguide modes with total optical decay rate $\Gamma$. Conservation of momentum requires that the emitter experience a momentum recoil opposite to the direction of photon emission.
The jump operators that describe the process of decay and momentum recoil within the context of the master equation, Eq. (\ref{eq:masterequation}), are
	\begin{align}
		\hat{L}_\ell &= \sqrt{\frac{\Gamma}{2}} \hat{\sigma}_- e^{ i \eta \hat{x} } \label{singlewaveguide1},\\
		\hat{L}_r &= \sqrt{\frac{\Gamma}{2}} \hat{\sigma}_- e^{ -i \eta \hat{x} } \label{singlewaveguide2},
	\end{align}
where $\hat{\sigma}_- = \ketbra{g}{e}$ is the electronic lower operator.
The momentum recoil that accompanies optical emission is generated by the dimensionless position operator $\hat{x} = \hat{v} + \hat{v}\dg$ with a strength determined by the Lamb-Dicke parameter $\eta\equiv k_0 x_{zpm}$, where $k_0 = 2\pi/\lambda_0$ is the magnitude of the emitted photon's wavevector.

An emitter initially prepared in the excited electronic state and motional ground state, $\ket{\psi(t_0)} = \ket{e} \otimes \ket{0}_m$, will decay into the waveguide modes.  For times long compared to the optical decay rate, $t \gg 1/\Gamma$, the initial excitation decays entirely into the waveguide. The emitter's long-time state after the light has been traced out, $\hat{\rho}(t) = \ketbra{g}{g} \otimes \hat{\rho}_{m}(t)$, is given by the master equation. 
Typically, after the emitted light propagates away the reduced motional state $\hat{\rho}_m(t)$ is mixed due to entanglement between the outgoing photon and the motional degree of freedom. 
Nevertheless, mixed motional states can in principle exhibit nonclassicality; we seek such a physical situation.

\begin{figure}
\centering
\includegraphics[width=15cm]{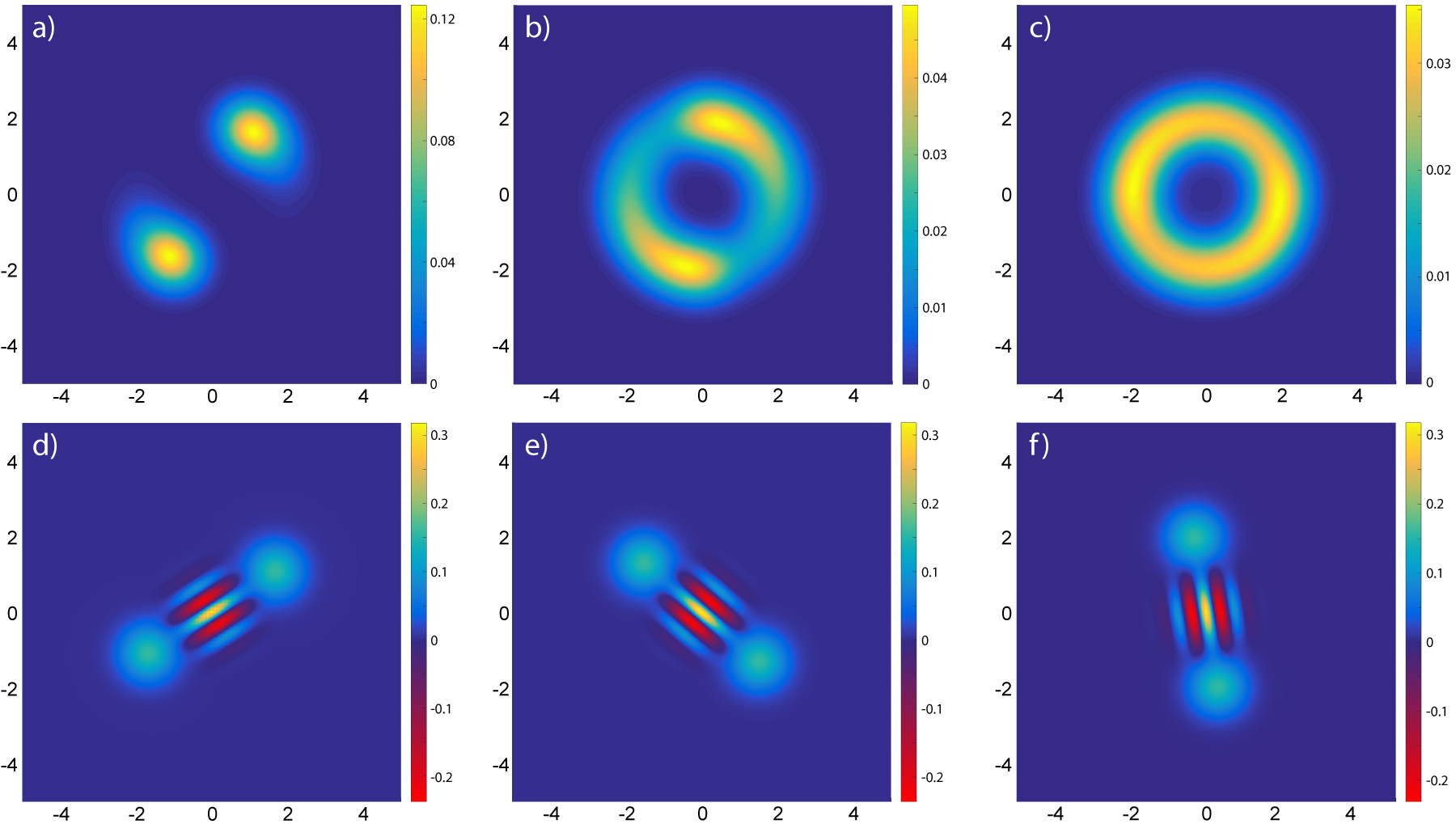}
\caption{Wigner functions for the reduced motional state of a trapped two-level emitter in the long-time limit (after decay to the ground electronic state). Negativity is indicated in red. Shown in a), b), and c) are unconditional Wigner functions for various optical decay rates $\Gamma/\omega_m = \{8,1,0.2\}$ and Lamb-Dicke parameter $\eta = 2$ when the photon field is traced out, see Fig. \ref{WaveguideSetup}(a). Fast decay with respective to the motional frequency yields a distinctly separated statistical mixture of momentum recoils as seen in a). For slower decay rates the recoils are smeared around the phase plane as seen in b) and c). Shown in d), e), and f) are conditional Wigner functions for the same parameter values when the waveguide modes are mixed on a beamsplitter the emitted photon is detected, see Fig. \ref{WaveguideSetup}b. In each, pure motional cat states are produced. The orientations are determined by the random photon detection times. }
\label{SinglewaveguideWigner}
\end{figure}

\begin{figure}
\centering
\includegraphics[width=15cm]{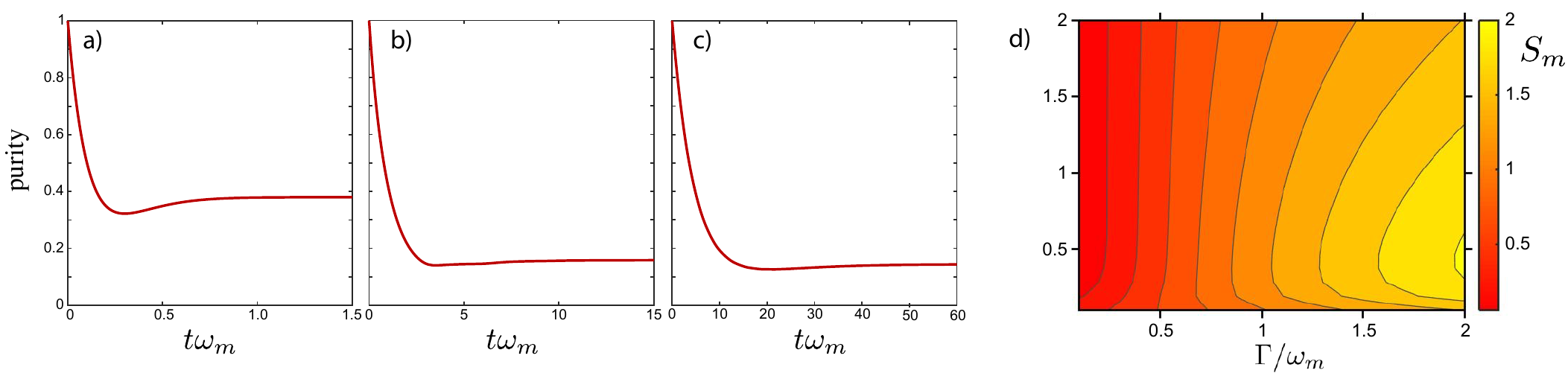}
\caption{Signatures of entanglement between the motional state and the emitted photon. Shown in a), b), and c) are the dynamics of the unconditional reduced motional purity, $\Tr [\hat{\rho}_m^2]$, during the approach to the long-time limit. The parameters values are the same as in Fig \ref{SinglewaveguideWigner}. Slower optical decay leads to lower purity as the momentum recoils are smeared in the phase plane. Shown in d) is the motional entropy, $S_m\equiv -{\rm Tr}[\hat{\rho}_m\ln \hat{\rho}_m]$, of the long-time motional state of the emitter as a function of the optical emission rate $\Gamma/\omega_m$ and Lamb-Dicke parameter $\eta$. The larger the value of $S_m$ the larger the correlation between the motional state and emitted optical field.}
\label{SinglewaveguideMotionalentropy}
\end{figure}

Decay into the left- and right-propagating waveguide modes results in both positive and negative displacements in momentum. These displacements are  correlated with the direction of the outgoing photon, resulting in a mixture of positive and negative momentum recoils when the photonic degrees of freedom are traced out. For an initially motionless trapped emitter we find two qualitatively different regimes.  If the coupling to the waveguide is large compared with the motional frequency $\Gamma/\omega_m > 1$, then the resulting displacement is an impulsive kick, which leads to long time states whose motional Wigner function exhibits two cleanly separated peaks, see Fig \ref{SinglewaveguideWigner}(a). As the coupling to the waveguide decreases, the decay and resulting motional displacement occur over a duration which is either commensurate with, or longer than, the motional period. The resulting motional Wigner function becomes smeared, see Figs \ref{SinglewaveguideWigner}(b) and \ref{SinglewaveguideWigner}(c). However, no parameter values $(\eta, \Gamma/\omega_m)$, were found that result in a long-time motional Wigner function with any negativity. That is, the atomic motion is left in a highly mixed ``classical'' motional state. The reduced motional state purity, $\Tr[\hat{\rho}_m^2(t)]$, is shown in \ref{SinglewaveguideMotionalentropy}. To further investigate the level of correlation between the motion and emitted optical fields we compute the motional entropy $S_m\equiv -{\rm Tr}[\hat{\rho}_m\ln \hat{\rho}_m]$, as a function of $(\eta, \Gamma/\omega_m)$. Since the joint state of the motion-excitation-optical field is pure at all times, and in the long-time limit the emitter is in the electronic ground state, the motional entropy is an entanglement measure for $t \rightarrow \infty$. Large motional entropy indicates a large correlation between the motion and the outgoing photon field.  From Fig. \ref{SinglewaveguideMotionalentropy}, we observe that the motion is heavily entangled with the outgoing photon pulses, preventing the reduced motional state from displaying negativity in its Wigner function.  

Information about the direction of momentum recoil contained in the outgoing photon field can be retrieved by measuring it. A ``click" at a photo-detector placed at each end of the waveguide determines whether the photon was emitted to the left or right. The corresponding recoil is simply a single coherent displacement of the momentum to the right or the left, respectively. In order to see nonclassical behavior, the ``which-way" information in the outgoing photon field must be coherently erased. This can be achieved by placing a beamsplitter before the detectors as depicted in Fig. \ref{WaveguideSetup}b. A photon exiting either port of the beamsplitter is no longer correlated with a single momentum kick but rather a superposition of recoils to the right and left. A click at either detector thus heralds a  motional state in a superposition of momentum recoils. Additionally, detection of the photon disentangles it from emitter, and resulting conditional motional state is pure. 

To investigate the \emph{conditional} motional state, we simulated the evolution using a stochastic master equation for photon counting \cite{GardZoll04}, which gives the emitter's state while both waveguide modes are continuously monitored. As long as only vacuum is measured at both detectors, the joint electronic-motional state remains in the initial state, $\ket{e} \otimes \ket{0}_m$, according to the quantum Zeno effect. When the photon is detected, the electronic state is projected into $\ket{g}$, and the motional state receives a superposition of momentum kicks with a relative phase that depends on which of the two detectors clicks. Examples of the resulting conditional Wigner functions are shown in Fig. \ref{SinglewaveguideWigner}(d)-(f), revealing near perfect motional cat states. 

It is well known that one can harness the nonunitarity of quantum measurements to generate nonclassical states. However, in practice the quality of the measurement has a large effect on the nonclassicality. Meanwhile  measurements at the quantum limit have become common but remain difficult to engineer. Thus, informed by the simple conditional model above where eliminating ``which-way" information is the key to nonclassicality, we now focus our attention on more elaborate settings where nonclassicality can arise even for \emph{unconditional} dynamics.

\section{Emitter in front of a mirror} \label{Sec:mirror}

\begin{figure}
\centering
\includegraphics[width=15cm]{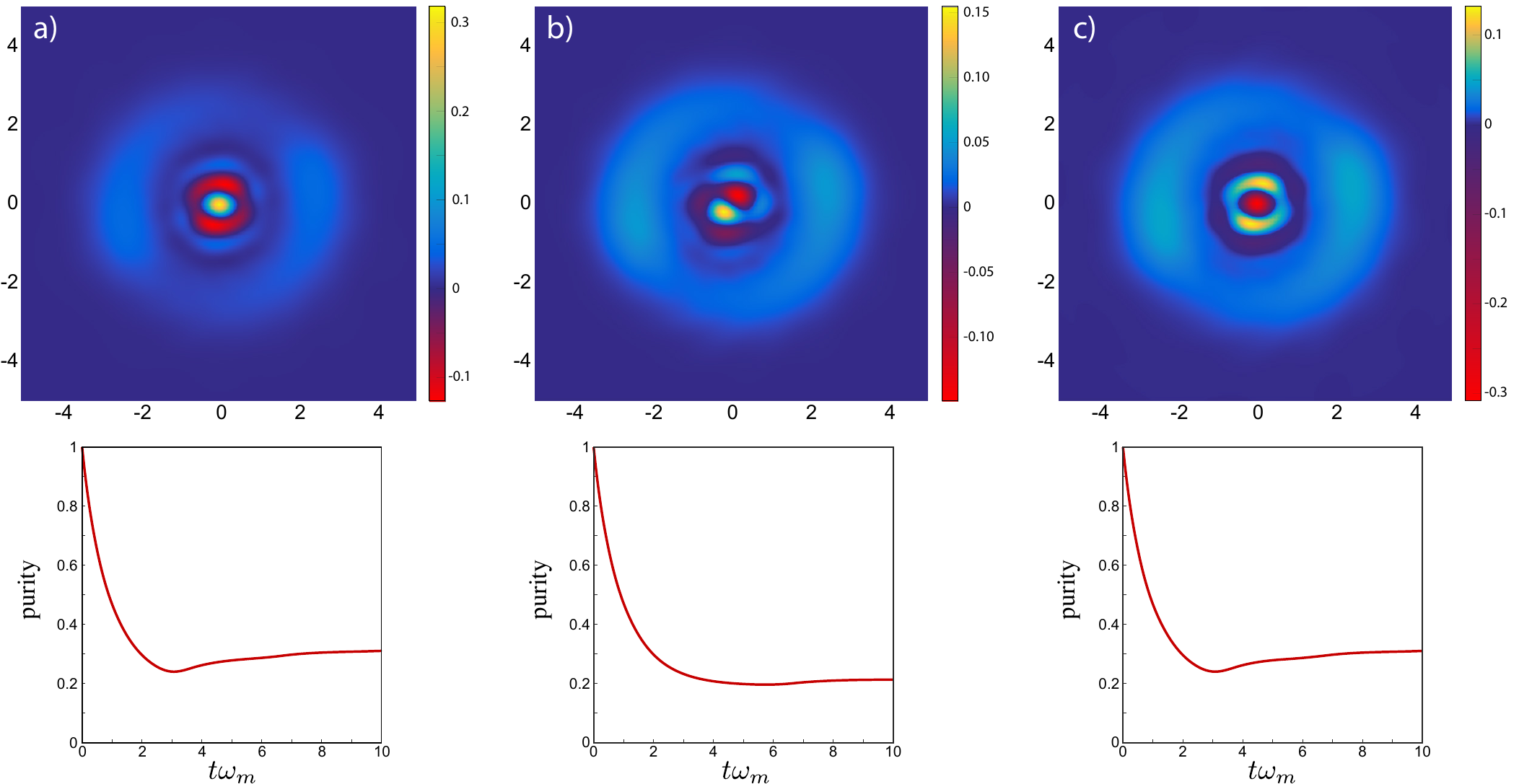}
\caption{Wigner functions for the reduced motional state of a trapped two-level emitter near a mirror on a waveguide in the long-time limit (after decay to the ground electronic state). Negativity is indicated in red. Shown in a), b), and c) are unconditional Wigner functions for the emitter placed at various distances from the mirror, parameterized by the propagation phases $\phi = \{0,\pi/4,\pi/2\}$ with Lamb-Dicke parameter $\eta = 2$. Below each Wigner function is the time evolution of the motional-state purity during the approach to the long-time limit. }
\label{fig:mirror1}
\end{figure} 

As we have seen above, nonclassical motional states can arise when ``which-way" information in the emitted photon is erased coherently. We first consider a straightforward way to achieve this by placing a perfect mirror near the trapped emitter to ensure that all emitted photons eventually exit in one direction down the waveguide, without relying on a conditional detection scheme. In the Lamb-Dicke regime, $\eta \ll 1$, motional coherence after spontaneous emission near a mirror has been observed \cite{Tomkovic:2011aa}.

To simulate the master equation for this situation, we use the SLH-formalism for quantum networks \cite{GougJame09a, GougJame09}, to find the appropriate Hamiltonian and jump operators. This formalism was designed to study quantum networks in the context of input-output theory and gives methods for reducing a complex set of interconnected network elements to a simpler description. Following the procedure detailed in \ref{AppendixA} we obtain the operators,
	\begin{eqnarray} 
		\hat{L}_\ell &=& \sqrt{2\Gamma}\,\hat{\sigma}_- \cos(\eta \hat{x}-\phi) \label{Mirror2} ,\\
		\hat{L}_r &=& 0 ,\\
		\hat{H}_{\rm sys}&=& \omega_m \hat{v}\dg \hat{v} + \frac{\Gamma}{2} |e\rangle\langle e|\sin(2\phi-2 \eta \hat{x}) ,\label{Mirror3}
	\end{eqnarray}
which are used in the master equation, Eq. (\ref{eq:masterequation}).
Here $\phi = 2 \pi d/\lambda_0$ is the phase acquired by the field as it propagates the distance $d$ between the emitter and mirror. The quantized position operator of the emitter, $\hat{x}=\hat{v}+\hat{v}^\dagger$, appears in an oscillatory term in both the jump operator, Eq. (\ref{Mirror2}), and the effective Hamiltonian. The latter describes an energy shift due to interaction with the reflected field, Eq. (\ref{Mirror3}). 

We again consider the decay of an initially excited emitter in the ground motional state, $\ket{\psi(t_0)} = \ket{e} \otimes \ket{0}_m$. The time evolution of the unconditional motional state of the emitter is found from the master equation in Eq. (\ref{eq:masterequation}) with the jump operators and Hamiltonian in Eqs. (\ref{Mirror2}) and (\ref{Mirror3}). We plot the Wigner functions for the long-time motional state in Fig. \ref{fig:mirror1}, for the emitter located at three positions relative to the mirror given by $\phi = \{0, \pi/4, \pi/2\}$. We observe a drastic difference from the previous case of emission into the bidirectional waveguide in Sec. \ref{sec:waveguide}. In all situations we observe that the motional state, though highly mixed, is also highly nonclassical due to significant negativity in the Wigner function. We thus have confirmed that if one arranges an optical setup where one forces the emitted light to re-interact with the emitter then the motional state can become nonclassical. 

In the absence of motional recoil ($\eta = 0$), the optical decay can be modulated by placing the emitter at different positions with respect to the mirror. When the emitter is placed at a node, $\phi = (2n+1)\pi/2$ for integer $n$, the reflected and emitted fields destructively interfere and  optical decay is entirely suppressed. This effect has been studied experimentally in free space by Blatt \emph{et al.} \cite{Eschner2001, Wilson2003, Bushev2004, Dubin2007}, who examined the position-dependent modulation of the fluorescence of a single ion positioned in front of a mirror. 

\begin{figure}
\centering
\includegraphics[width=15cm]{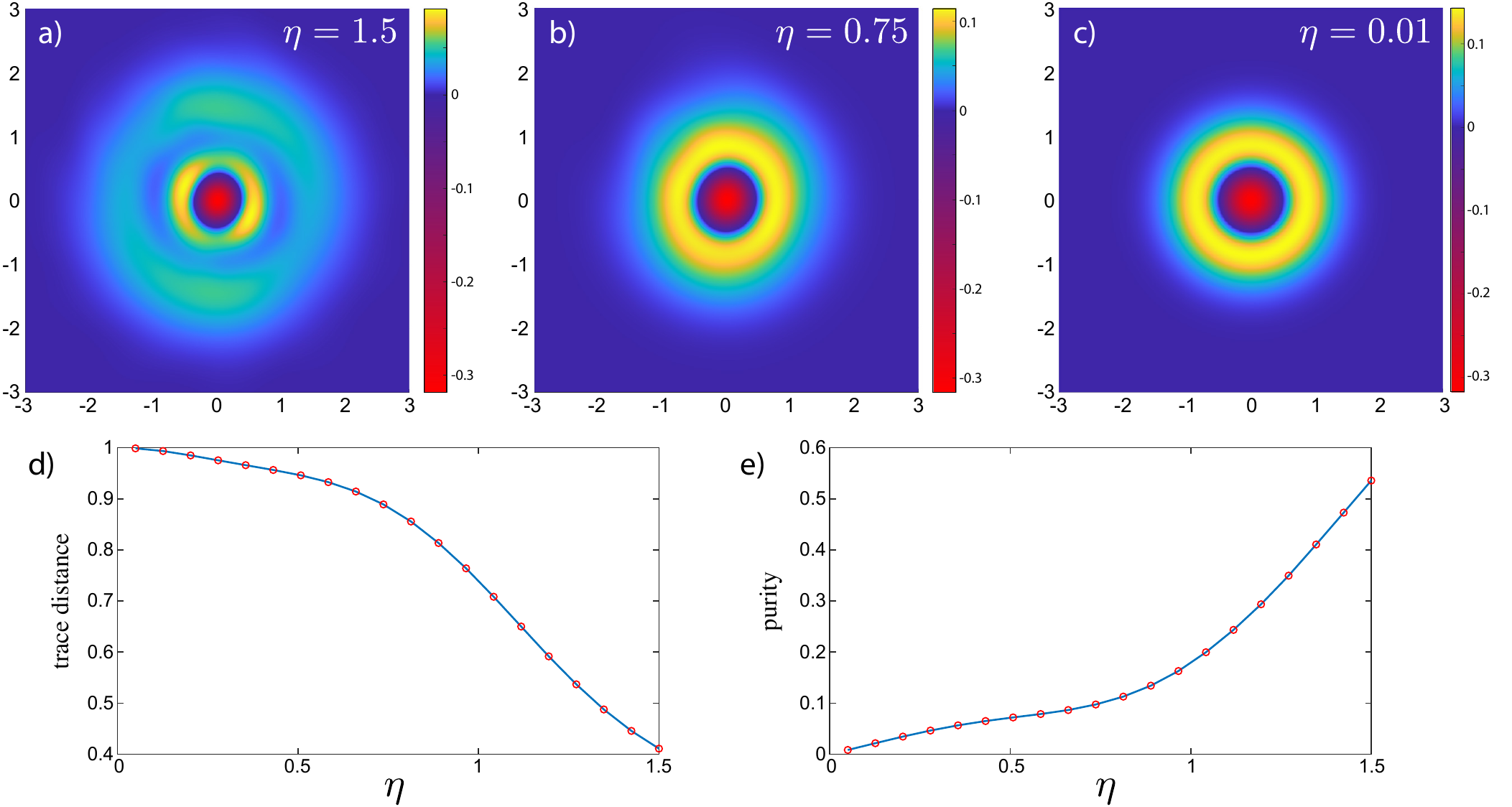}
\caption{Approach to a single-excitation Fock state in the motional degree of freedom as the Lamb-Dicke parameter $\eta$ is reduced. Shown are Wigner functions for the long-time motional state for an emitter near a mirror placed at a node, $\phi = \pi/2$. In (a), (b), and (c) the Lamb-Dicke parameter is $\eta = \{1.5, 0.75, 0.01\}$, respectively. In all plots the optical decay is $\Gamma/\omega_m = 1/4$. (d) Purity of the long-time motional state as a function of $\eta$. (e) Trace distance between the long-time motional state and a pure, single-excitation Fock state of motion. }
\label{mirror_Fock}
\end{figure}

For nonzero Lamb-Dicke parameter, $\eta > 0$, the electronic and motional degrees of freedom are coupled, evident in the jump operator Eq. (\ref{Mirror2}). The extent of the emitter's spatial wave function beyond the node, even when in the motional ground state, breaks the perfect destructive interference and allows decay in to the waveguide. This effect can be used to engineer pure, single-excitation Fock states of motion $\ket{1}_m$. The emitter is placed at a node, $\phi=\pi/2$, and the Lamb-Dicke parameter $\eta$ is varied, see Fig. (\ref{mirror_Fock}). As $\eta \rightarrow 0$, the long-time motional state becomes a pure state, as seen in Fig. \ref{mirror_Fock}(d). To verify that the pure state is indeed a single-excitation Fock state of motion, $\ket{1}_m$, in Fig. \ref{mirror_Fock}(e) we plot the trace distance, $\frac{1}{2}\Tr[ \sqrt {(\hat{\rho}_m - \hat{\sigma})^2} ]$, 
where  $\hat{\rho}_m$ is the long-time motional state and $\hat{\sigma} = \ketbra{1}{1}_m$. Again, as $\eta \rightarrow 0$, the state asymptotically approaches $\ket{1}_m$.

\section{Emitter coupled to a waveguide via a toroidal cavity} \label{sec:toroidal}

\begin{figure}
\centering
\includegraphics[width=7.5cm]{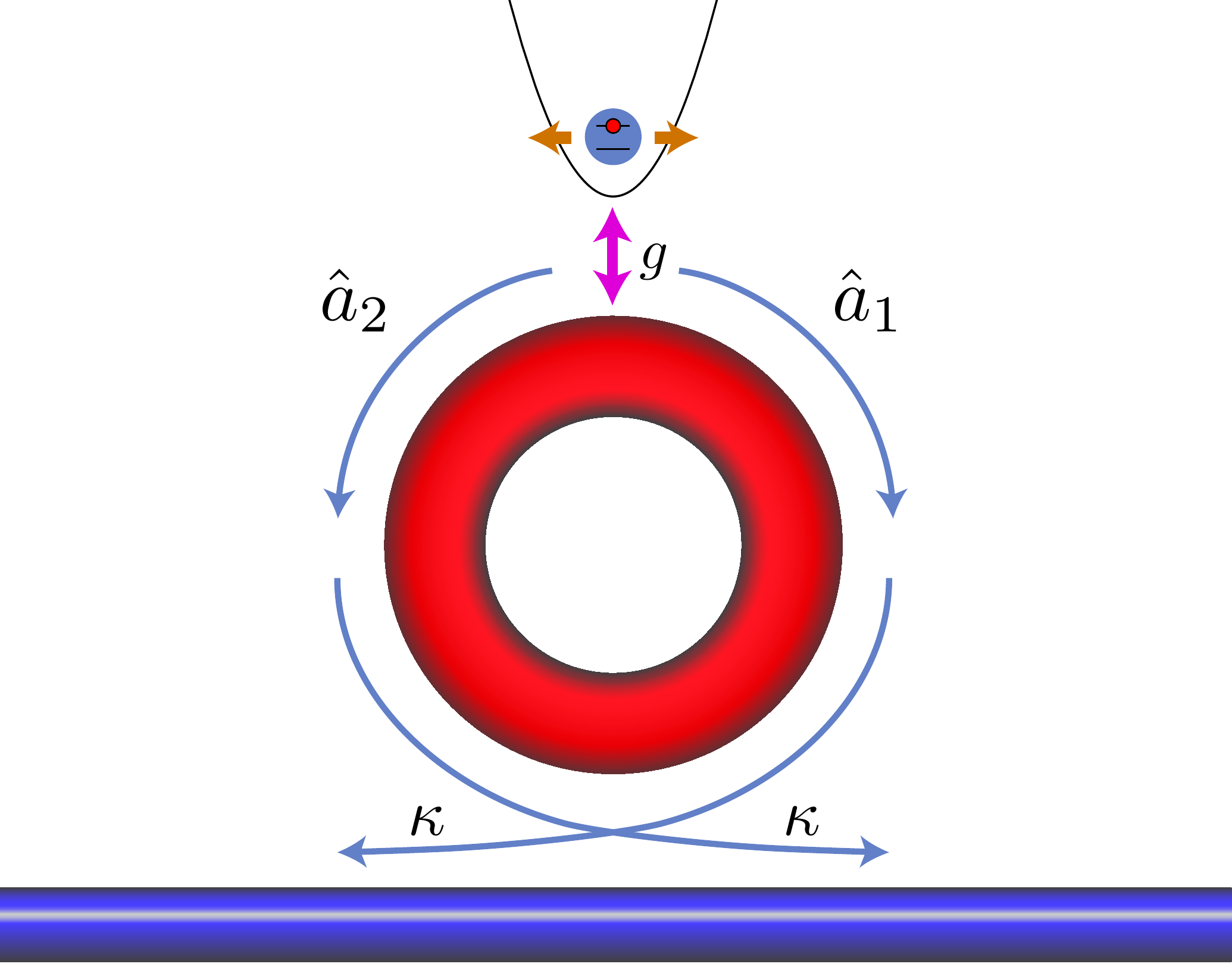}
\caption{Schematic setup for a harmonically trapped two-level emitter coupled symmetrically with strength $g$ to degenerate clockwise ($\hat{a}_1$) and anticlockwise ($\hat{a}_2$) circulating optical modes of a toroidal resonator. The resonator modes are outcoupled with strength $\kappa$ to the left- and right-propagating modes of a 1D waveguide. The emitter is detuned by $\Delta$ from the toroidal cavity modes.}
\label{ToroidalSetup}
\end{figure}

We now consider a physical architecture that has been experimentally realized with tapered optical nanofibers by Rauschenbeutel and co-workers \cite{Sayrin2015, Scheucher2016}.
A trapped atom is strongly coupled to circulating optical modes in a ``bottle resonator'' fabricated in an optical fiber, which is evanescently in/out coupled to a tapered nanofiber. 
We model this architecture as depicted in Fig. \ref{ToroidalSetup}. The trapped emitter couples with strength $g$ to clockwise(counterclockwise) circulating photons in a toroidal resonator with frequency $\omega_R$ and annihilation operators $\hat{a}_1(\hat{a}_2)$. The emitter is detuned from the resonance frequency of the toroidal resonator modes by $\Delta \equiv \omega_
R -\omega_0$, and the respective resonator modes are coupled to the left- and right-propagating waveguide modes with rate $\kappa$. Rather than breaking the directional symmetry as in Section \ref{Sec:mirror}, here the left- and right-propagating waveguide modes are treated on equal footing. The toroidal resonator facilitates a photon reinteracting with the emitter, which can then re-emit into either resonator mode, effectively scrambling the `which way' information. 
In the interaction picture the Hamiltonian governing the coupling between the emitter and toroidal resonator modes is
\begin{eqnarray}
\hat{H}_{\rm sys} &=& -\frac{\Delta}{2}\hat{\sigma}_z+\omega_m\hat{v}^\dagger\hat{v}+ g \left[\hat{\sigma}_-(\hat{a}_1^\dagger e^{i\eta\hat{x}}+\hat{a}_2^\dagger e^{-i\eta\hat{x}})+{\rm h.c.}\right]\label{Toroid3}, 
\end{eqnarray}
and the jump operators describing coupling from the toroidal resonator to the waveguide modes are,
\begin{eqnarray}
\hat{L}_\ell &=& \sqrt{\kappa}\,\hat{a}_1\label{Toroid1} \\
\hat{L}_r &=& \sqrt{\kappa}\,\hat{a}_2\label{Toroid2} .
\end{eqnarray}

\begin{figure}
\centering
\includegraphics[width=15cm]{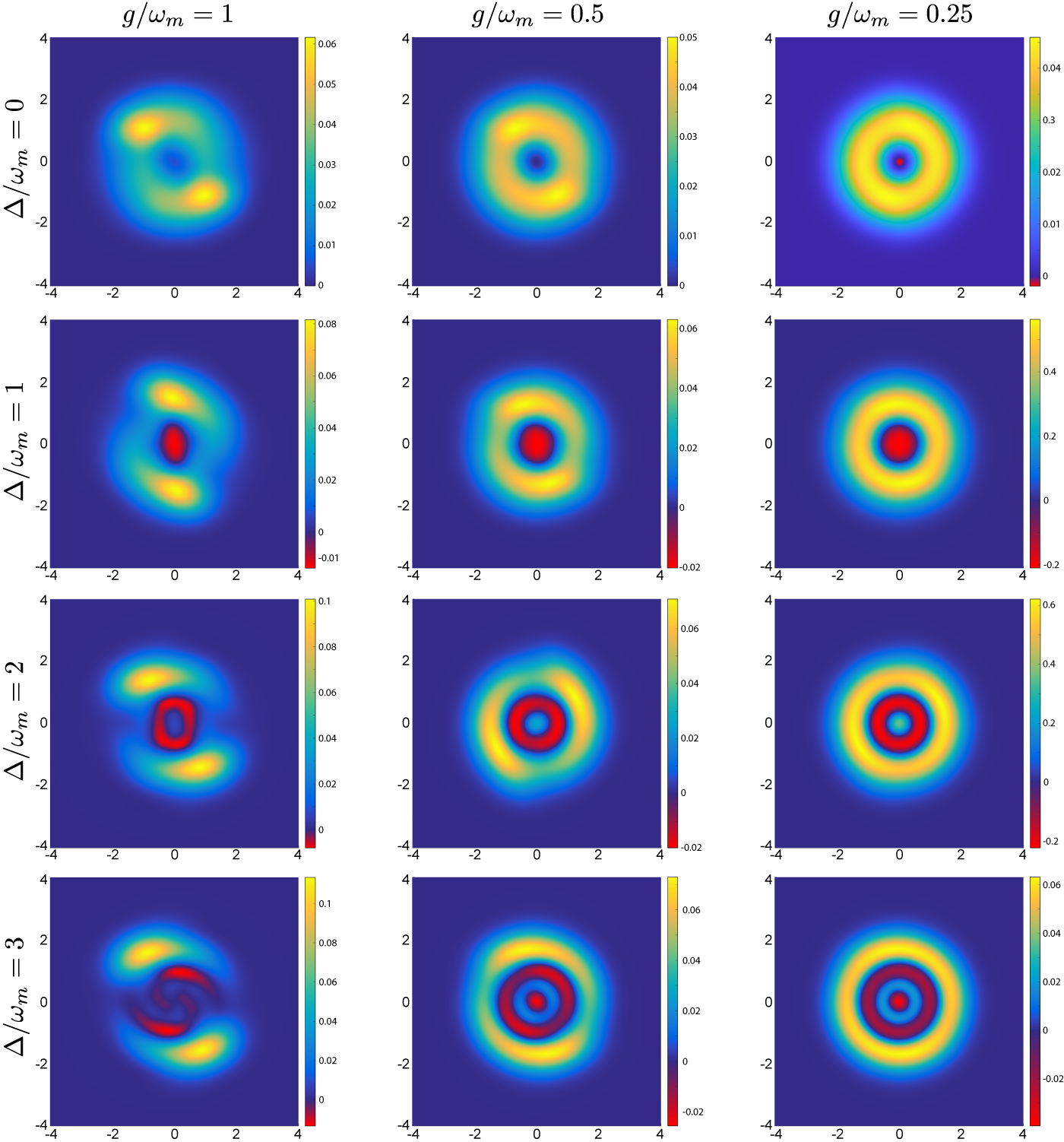}
\caption{Wigner functions for the reduced motional state for a trapped two-level emitter coupled to a toroidal resonator in the long-time limit (after decay to the ground electronic state). Negativity is indicated in red. The columns have fixed coupling rates between the emitter and resonator modes $g/\omega_m = \{1,0.5,0.25\}$, and the rows have fixed emitter-resonator detunings $\Delta/\omega_m = \{0,1,2,3\}$. For all simulations the cavity decay rate and Lamb-Dicke parameter are fixed: $\kappa = 2$, $\eta = 2$. }
\label{toroidal_wig}
\end{figure} 

\begin{figure}
\centering
\includegraphics[width=15cm]{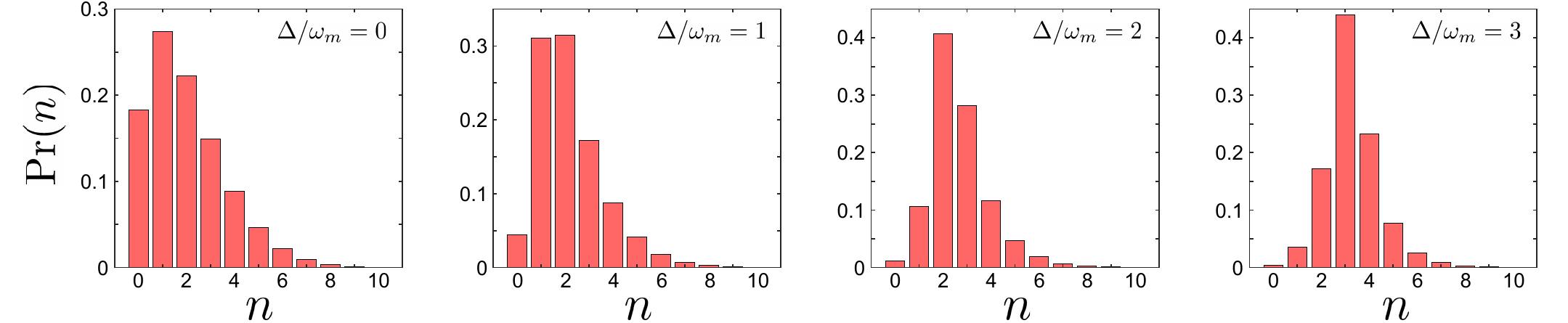}
\caption{Number distribution of the density matrix for the reduced motional states in the rightmost column  of Fig. \ref{toroidal_wig} ($g/\omega_m = 0.25$). }
\label{toroidal_numberdistributions}
\end{figure} 

\begin{figure}
\centering
\includegraphics[width=15cm]{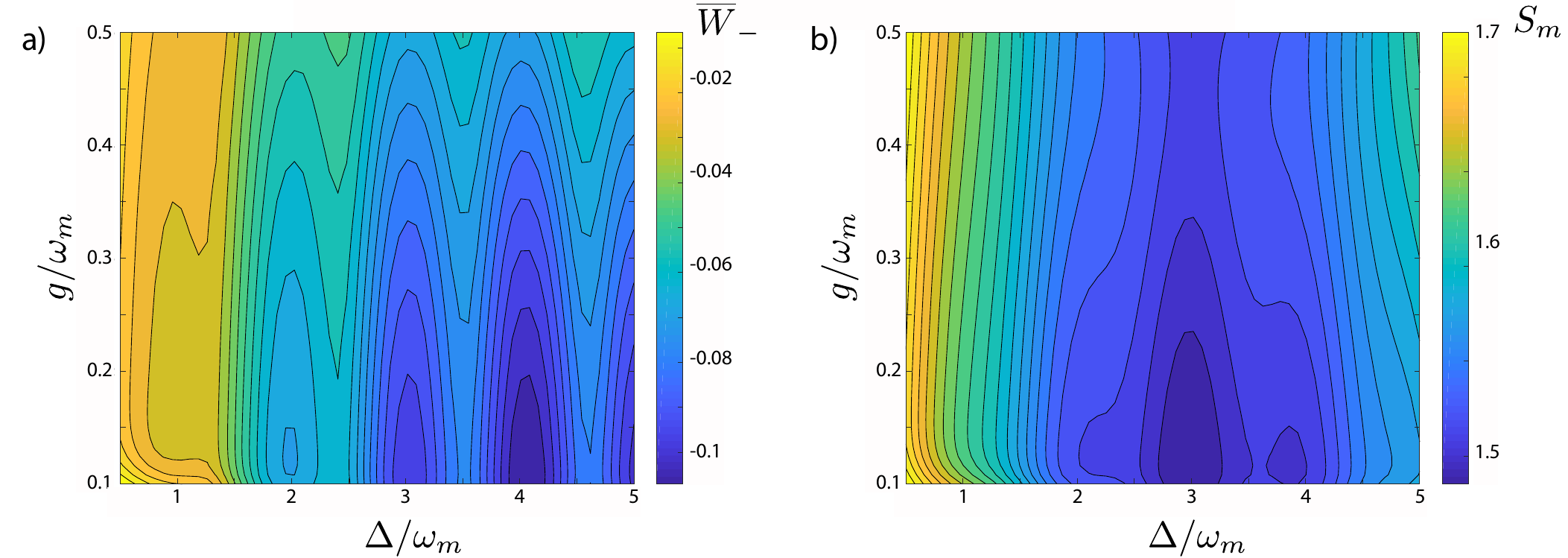}
\caption{Properties of the long-time motional state. (a) Integrated negativity of the Wigner function, Eq. (\ref{netwig}). (b) Motional entropy. Fixed parameters are $\kappa = 2$, $\eta = 2$. }
\label{toroidal_scans}
\end{figure} 

We initialize the system with the emitter in the excited electronic state and ground motional state, and the toroidal cavities in vacuum. The open-system dynamics of the joint emitter-resonator system are given by the master equation in Eq (\ref{eq:masterequation}) with the Hamiltonian and jump operators above.  In the long-time limit we examine the reduced motional state, $\hat{\rho}_m$, by tracing out the electronic and resonator degrees of freedom after the emitter has decayed to the electronic ground state and all the light has exited into the waveguide. Example Wigner functions for $\hat{\rho}_m$ are show in Fig. (\ref{toroidal_wig}) for fixed Lamb-Dicke parameter $\eta = 2$ and cavity decay $\kappa = 2$. Comparing the columns reveals that decreasing the emitter-resonantor coupling with respect to the motional frequency, $g/\omega_m$, eliminates the phase information. For small $g/\omega_m$ the motional state performs many oscillations during a Jaynes-Cummings-type transfer of excitation from emitter to resonator.  

Comparing the rightmost rows of Fig. \ref{toroidal_wig} ($g/\omega_m = 0.25$) shows that increasing the emitter-resonator detuning produces rings of negativity in the Wigner function similar to those for Fock states of increasing $n$. However, the motional states cannot be a  Fock state since they are highly mixed. Indeed, the motional states are in fact almost perfectly diagonal in the Fock basis, indicating a mixture over perfect Fock states. The number distributions for the diagonal elements of $\hat{\rho}_m$ are shown in Fig. (\ref{toroidal_numberdistributions}). As the detuning is increased, the relative amount of vacuum, $n=0$, diminishes, and the resulting motional states are mixtures of nonclassical Fock states.

To compare nonclassicality of the long-time motional states, we compute the \emph{integrated negativity} of the motional Wigner function,
\begin{equation}
\overline{W}_-\equiv \int_{A_-}\, W(x,p)\,dxdp \;\;,
\label{netwig}
\end{equation}
where $A_-$ are the areas in phase space in which the motional Wigner function is negative. We seek parameter values that leave $\overline{W}_-$ as negative as possible indicating a highly nonclassical motional state. In Fig \ref{toroidal_scans}(a), we observe that nonclassical states require a small coupling between the emitter and the cavities (so that the light circulating in the toroidal cavities can interact with the emitter for longer), and with a periodic dependence on the detuning.

\begin{figure}
\centering
\includegraphics[width=15cm]{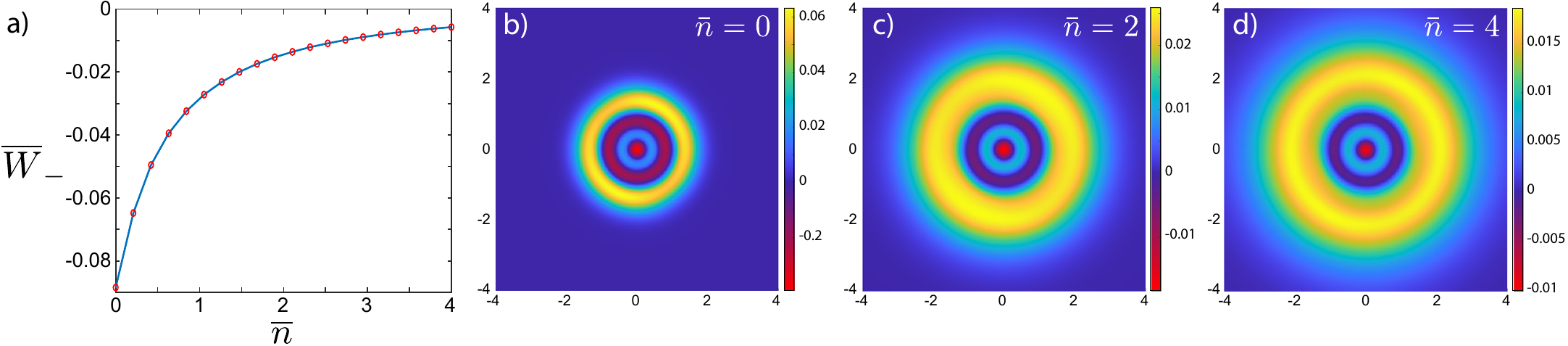}
\caption{(a) Integrated negativity of the long-time motional state as the initial thermal occupation, $\bar{n}$, is increased. Long-time motional state given a thermally distributed initial state. (b), (c), and (d) Wigner functions for several values of initial thermal occupation. The simulation parameters are $g/\omega_m = 0.25$, $\Delta/\omega_m = 3$, $\kappa = 2$, $\eta = 2$.  }
\label{MEWignerThermal}
\end{figure}

Finally, nonclassicality of the long-time motional state has some robustness to situations where the trapped emitter is not initially in the motional ground state. This is common in realistic physical settings where perfect ground-state cooling is not achieved. As shown in Fig. \ref{MEWignerThermal}, integrated negativity persists when the emitter's motional degree of freedom is initially in a thermal state with mean occupation number $\bar{n} \equiv {\rm Tr}[ \hat{\rho}_m(0) \hat{v}^\dagger\hat{v}]$. 

\section{Conclusion}

We have investigated the production of nonclassical motional states in several physical architectures based on coupling a trapped two-level emitter to a 1D photonic waveguide. In each case the fundamental optomechanical coupling between the internal degree of freedom and the motional degree of freedom is provided by momentum recoil as a photon is emitted, as required by conservation of momentum. While nonclassical states can be readily produced by detecting the photon, we have shown that even unconditional motional states can exhibit significant nonclassical features in the long-time limit. Our method for the production of nonclassical motional states relies only on the fundamental coupling between an emitter and its electromagnetic environment making it relatively straightforward compared to procedures involving complicated measurement and feedback. 
An interesting extension to generate larger, potentially entangled motional states involves multiple emitters in a single trap next to a waveguide. Cooperative emission might yield a superradiant recoil \cite{Damanet2016} that can be harnessed to produce extremely large motional nonclassical motional states.

\section{Acknowledgements}
B.Q.B. and J.T. acknowledge support from the Australian Research Council Centre of Excellence in Engineered Quantum Systems CE110001013.
B.Q.B. acknowledge support from the ARC Centre of Excellence for Quantum Computation and Communication Technology (Project No.\ CE170100012).

\section{References}

\providecommand{\newblock}{}

\appendix

\section{Emitter trapped in a Fabry-Perot setup: SLH formalism} \label{AppendixA}
\begin{figure}
\centering
\setlength{\unitlength}{1cm}
\begin{picture}(10,14.2)
\put(0,7.5){\includegraphics[width=11.5cm]{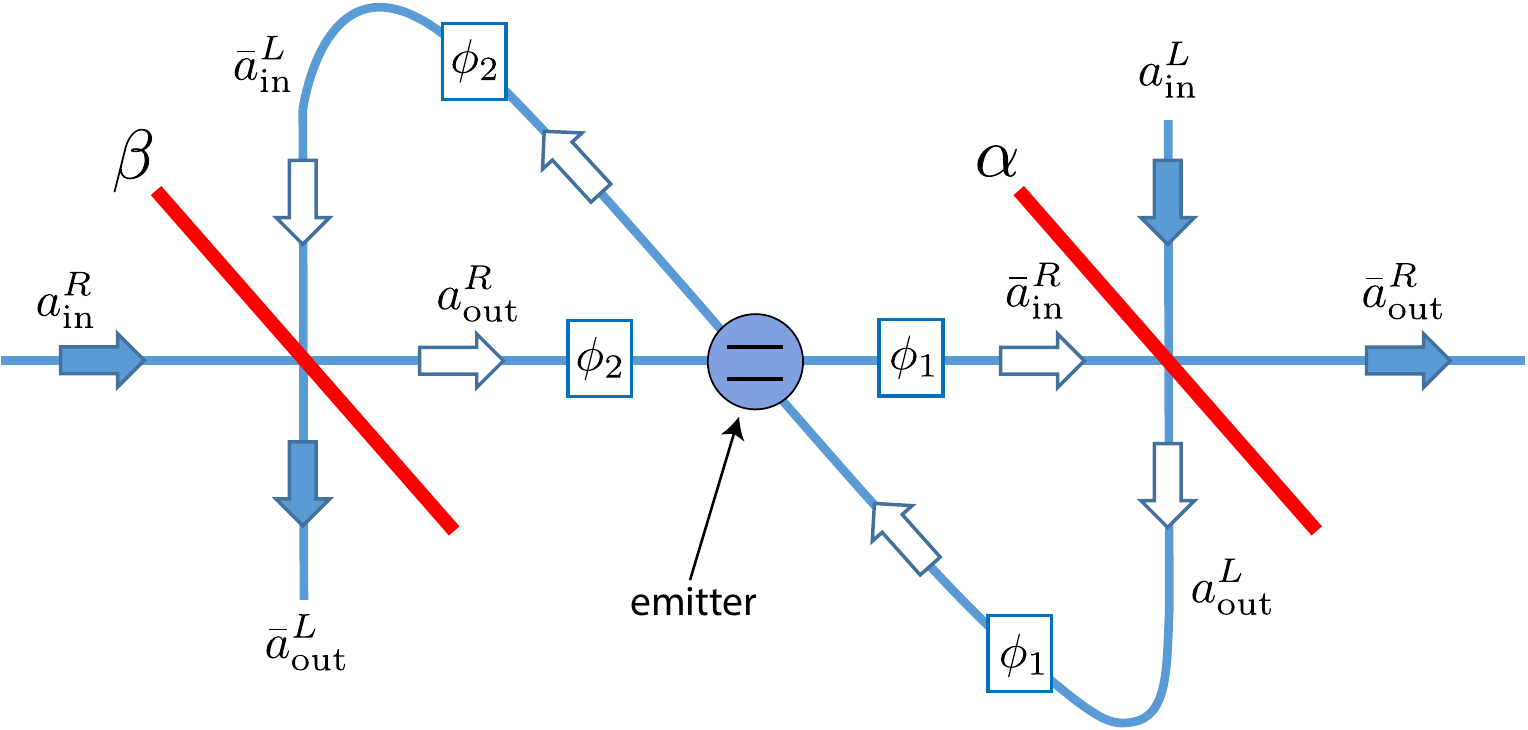}}
\put(2,0){\includegraphics[width=8.5cm]{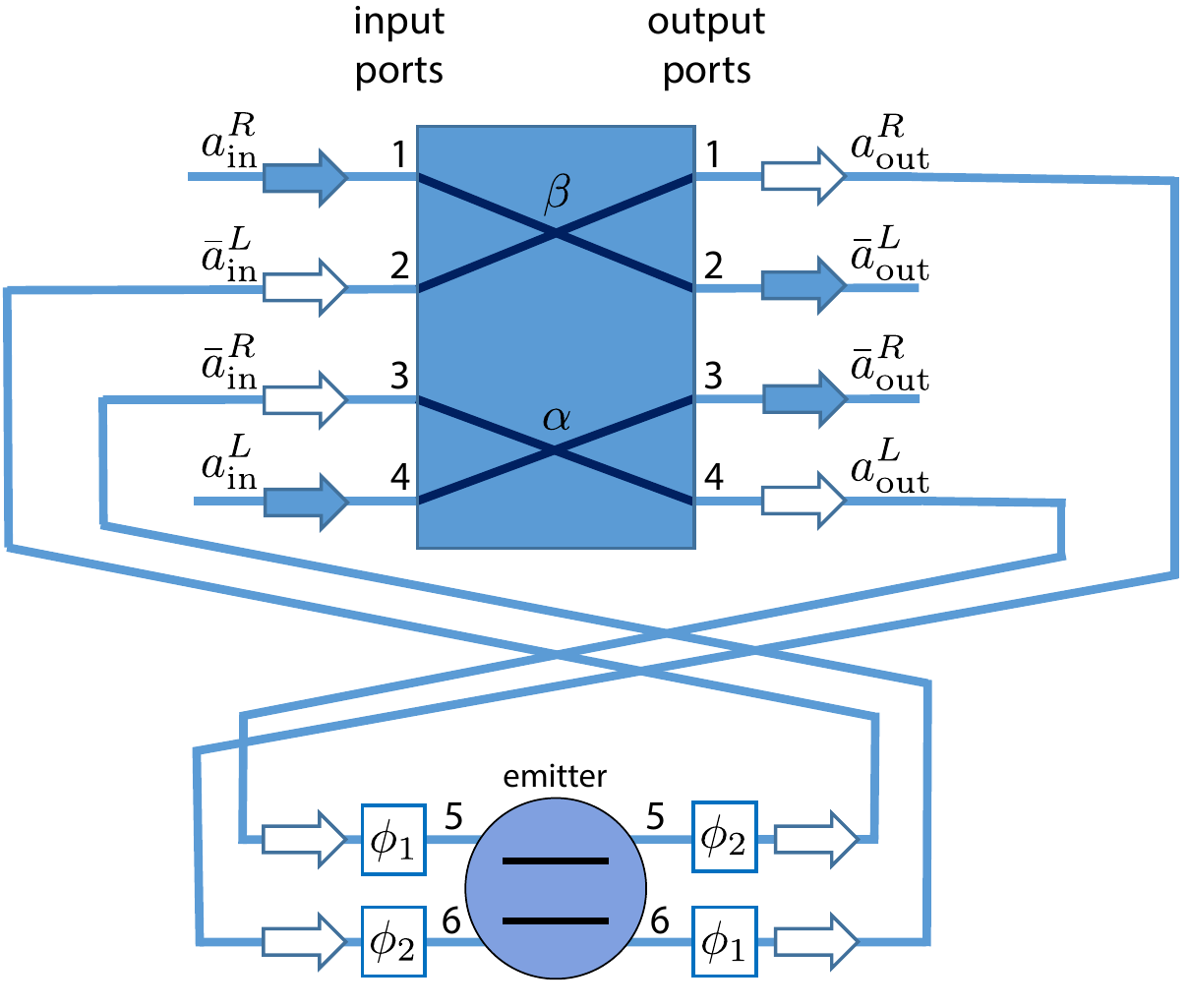}}
\put(0,12.5){a)}
\put(0,5.5){b)}
\end{picture}
\caption{Optical network for a two-level emitter coupled to a 1D waveguide with an embedded Fabry-Perot cavity, modeled as two interconnected beamsplitters. a) Graphical schematic of the emitter in between two beamsplitters. The left(right) beamsplitter has reflectivity $\beta(\alpha)$. The straight horizontal optical path depicts the right-propagating waveguide mode coming in from the left while the curved optical path depicts the left-propagating mode coming in from the right. Solid arrows indicate external inputs and output, while outlined arrows indicate internal connections. b) Abstract network diagram of the same setup where we label the six input and output ports of the beamsplitters and emitter. Four internal interconnects link the internal inputs and outputs that are eliminated in the feedback reduction.}
\label{FPSetup}
\end{figure}

Here we derive the jump operators and Hamiltonian used in the master equation, Eq. (\ref{eq:masterequation}), for a trapped two-level emitter situated between two partially reflecting mirrors along a 1D waveguide, which we call a Fabry-Perot (FP) setup. 
The mirrors serve to reroute the outgoing photon field from the emitter to re-interact with the emitter, thereby erasing the ``which-way" information. To model such a system we invoke the theory of cascaded quantum systems for input-output theory. The schematic setup is shown in Fig \ref{FPSetup}a, where the left- and right-moving input modes, labeled as $a_{in}^L\,(a_{in}^R)$, enter the FP setup from the right and left sides, respectively. Each mirror in the FP setup is modeled as a beamsplitter that couples the waveguide modes on either side, and the emitter is placed between them. 

The input modes each interact with a beamsplitter with reflectivity $\alpha(\beta)$, resulting in scattered output fields that are transmitted in mode labeled $a_{out}^L\,(a_{out}^R)$, or reflected into mode $\bar{a}_{out}^R\,(\bar{a}_{out}^L)$. From each beamsplitter the transmitted mode then propagates to the emitter and is absorbed and reemitted in both directions. As the fields propagatesbetween the beamsplitters and the emitter they acquire propagation phases $\phi_1$ and $\phi_2$, respectively. This FP setup can be modelled as an input-output network with interconnected linkages from each of the three components (two BS and the emitter).  The SLH formalism, reviewed in Ref. \cite{Combes2016}, was designed to model such networks. Within the formalism each component is fully described by three objects: a unitary scattering matrix {\bf S}, a vector of jump operators $\bf L$, and a Hamiltonian $\hat{H}$. Collected together as $G = ({\bf S}, {\bf L}, \hat{H} )$, these are referred to as an SLH-triple. We leave operator hats off of {\bf S} and {\bf L} for notational convenience. 

When multiple components are interconnected, the formalism provides rules to collapse the internal connections such that the entire network can itself be described by a single SLH-triple \cite{GougJame09a,GougJame09,Combes2016,CookSchuClel18}. 
Specifically, Rule 4 of Ref. \cite{Combes2016} describes how to perform \emph{feedback reduction} of an SLH-triple by eliminating an internal interconnect. Once all the interconnects have been eliminated, the unconditional dynamics of the network as a whole are given by a master equation, Eq. (\ref{eq:masterequation}), using the resulting ${\bf L}$ and $\hat{H}$. Below we will discuss a particular caveat associated with feedback reduction of optical quantum networks; at the moment we focus on a short review of the procedure within the context of our problem. 

Before applying feedback reductions we recast the setup in Fig \ref{FPSetup}a, into the more abstract form depicted in Fig \ref{FPSetup}b. We label the input ports from $n_{in}=1,2,\dots 6$, and output ports from $n_{out}=1, 2, \dots 6$. We will have to perform four reductions as the network at large has two physical inputs and two outputs (arrows in Fig \ref{FPSetup}b). 
Before reductions, the SLH-triple for this six-port input-output device operates on input operator vector $\vec{x}=(a_{in}^R, \bar{a}_{in}^L, a_{in}^L,\bar{a}_{in}^R, emitter_{L}, emitter_{R})$, and gives the output vector $\vec{y}=(a_{out}^R, \bar{a}_{out}^L, a_{out}^L, \bar{a}_{out}^R, emitter_L, emitter_R)$, where $emitter_{R/L}$ denotes the emitter coupling to the right- and left-moving modes. Using this ordering we denote the scattering matrix for the setup as
\begin{align}	
		\mathbf{S} = 
		\left(
		\begin{array}{ccc}
			\mathbf{S}^L_{\rm BS}(\beta) & \mathbf{0} & \mathbf{0}\\
			\mathbf{0} & \mathbf{S}^R_{\rm BS}(\alpha) & \mathbf{0} \\
			\mathbf{0} & \mathbf{0} & \mathbf{S}_{\rm emitter}
		\end{array}
		\right), 
	\end{align}
where $\mathbf{S}^{L\{R\}}_{\rm BS}(\alpha\{\beta\})$ is the two-port scattering matrix for the left$\{$right$\}$ beamsplitter. The full $\mathbf{S}$ matrix is fully represented as 
\begin{equation}
\mathbf{S}=\left(\begin{array}{cccccc}
iC\beta & S\beta & 0 & 0 & 0 & 0 \\
S\beta & iC\beta & 0 & 0 & 0 & 0\\
0 & 0 & iC\alpha & S\alpha & 0 & 0\\
0 & 0 & S\alpha & iC\alpha & 0 & 0\\
0 & 0 & 0 & 0 & e^{i(\phi_1+\phi_2)} & 0 \\
0 & 0 & 0 & 0 & 0 & e^{i(\phi_1+\phi_2)}
\end{array}\right) ,
\label{bigS}
\end{equation}
where $C\beta\equiv \cos\beta$, $S\beta\equiv\sin\beta$, etc. The last two diagonal entries correspond to the phase acquired by the left- and right-moving modes as they propagate once between the beamsplitters. The vector of jump operators $\hat{\mathbf{L}}$ describes the linear coupling of the components to the input/output fields. As the beamsplitters have no internal dynamics, the only nontrivial $\hat{L}$ is that describing the emission of photons. Within the master equation, it drives spontaneous emission of the emitter (and recoil), into the left- and right-moving modes (taking into account the propagation phases). The vector of jump operators is then
\begin{equation}
\mathbf{L}=\sqrt{\frac{\Gamma}{2}}\,\hat{\sigma}_-\,\left(\begin{array}{c}
0\\0\\0\\0\\e^{i\phi_2 + ik \hat{x}}\\e^{i\phi_1 - ik \hat{x}}
\end{array}\right)
\label{bigL}.
\end{equation}
Finally, we work in the interaction picture with respect to the electronic degrees of freedom, so the Hamiltonian $\hat{H}$ is simply that of the harmonic motion in the trap $\hat{H}_{\rm sys}=\omega_m \hat{v}^\dagger\hat{v}$. 

We must now interconnect the various internal network links in the optical network in Fig \ref{FPSetup}b. Some details how this is achieved is outlined in  \ref{Reduction}. Once all the internal links are reduced one  obtains a two-input, two-output SLH-triple. Here we focus on two specific cases of interest, the first being when the left BS is completely transparent, e.g. there is no left BS, $\beta=0$, and the right BS is fully reflective, i.e. is a mirror, $\alpha=\pi/2$. The second case attempts to model the emitter in a Fabry-Perot cavity with 50:50 mirrors. To model this we set  each BS to be a 50:50 BS, e.g. $\beta=\alpha=\pi/4$. First, however, we derive the general SLH-triple for arbitrary $\alpha$, and $\beta$, but set the emitter at the mid-point between the BS e.g. ($\phi_1=\phi_2\equiv \phi$), for simplicity.
 After all feedback and internal link reductions have been performed, we arrive at the generalized final two-input, two-output SLH-triple: 
\begin{eqnarray} \label{eq:SLHFPsetup}
\mathbf{S} &=& \frac{1}{F_1}\,\left(\begin{array}{cc}
\sin(\beta)-e^{4i\phi}\sin(\alpha) & -e^{2i\phi}\cos(\alpha)\cos(\beta)\\
-e^{2i\phi}\cos(\alpha)\cos(\beta) & \sin(\alpha)-e^{4i\phi}\sin(\beta)
\end{array}\right)\label{FP1}\\[1em]
\mathbf{L} &=& \frac{\sqrt{\Gamma/2}\,\hat{\sigma}_-}{F_1}\,\left( \begin{array}{c}
\cos(\beta)[e^{-i(\phi-k\hat{x})} + e^{i(\phi-k\hat{x})}\sin(\alpha)]\\
\cos(\alpha)[e^{-i(\phi+k\hat{x})}+e^{i(\phi+k\hat{x})}\sin(\beta)]
\end{array}\right) \label{FP2}\\[1em]
\hat{H}_{\rm sys} &=& \omega_m \hat{v}\dg \hat{v}+|e\rangle\langle e| 
	\frac{\Gamma e^{4i\phi}}{2 F_1 F_2}\Big[
	\sin(\beta)(1+\sin^2(\alpha))\sin(2\phi+2k\hat{x})   \\
	&+&\sin(\alpha)(1+\sin^2(\beta))\sin(2\phi-2k\hat{x})+2
	\sin(\beta)\sin(\alpha)\sin(4\phi)\Big]\label{FP3} \nonumber
\end{eqnarray}
where $F_1=1-e^{4i\phi}\sin(\alpha)\sin(\beta)$, and $F_2=e^{4i\phi}-\sin(\alpha)\sin(\beta)$. Note that the scattering matrix is unitary: $\mathbf{S}\dg \mathbf{S} = \mathbf{S} \mathbf{S}\dg= \mathbf{I}$.
By setting $\beta=0$ and $\alpha=\pi/2$, we immediately obtain the SLH-triple presented in Sec. \ref{Sec:mirror}, for the emitter in front of a mirror. In this case $\phi$ represents the phase accrued by the light traveling between the emitter and the mirror.

\subsection{Emitter positioned midway between two 50/50 mirrors - a Fabry-Perot optical cavity}

Let us now consider the case where both beamsplitters are partially transmitting. In this case the emitted photon in either direction is partially reflected and partially transmitted, allowing it to make many roundtrips within the effective cavity---a trapped optical mode. While the setup is physically feasible, one must take care when applying the SLH formalism. It was recently pointed out and studied  by Tabak and Mabuchi \cite{Tabak2015}, that when trapped optical modes appear in quantum networks, in some cases they must be explicitly quantized. 

Instead, we continue without explicitly quantizing the trapped mode between the two BS in the Fabry-Perot cavity, in order to illustrate the problems that can arise. Using 50/50 beamsplitters ($\alpha = \beta = \pi/4$) the SLH-triple in Eq. (\ref{eq:SLHFPsetup}) becomes,
\begin{eqnarray}
\mathbf{S} &=& \frac{1}{\sqrt{2} (1- e^{4i\phi}/2)}\,\left(\begin{array}{cc}
1-e^{4i\phi} & -e^{2i\phi}/\sqrt{2}\\
-e^{2i\phi}/\sqrt{2} &1-e^{4i\phi}
\end{array}\right)\label{FP1}\\[1em]
\mathbf{L} &=& \sqrt{\frac{\Gamma}{2}} \frac{\hat{\sigma}_-}{\sqrt{2} (1- e^{4i\phi}/2)}\,\left( \begin{array}{c}
e^{-i(\phi-k\hat{x})}+e^{i(\phi-k\hat{x})}/\sqrt{2}\\
e^{-i(\phi+k\hat{x})}+e^{i(\phi+k\hat{x})}/\sqrt{2}
\end{array}\right) \label{FP2}\\[1em]
\hat{H}_{\rm sys} &=& \omega_m \hat{v}\dg \hat{v}  \\
&&+|e\rangle\langle e| 
\frac{\Gamma}{2} \frac{ 1 }{5 - 4 \cos 4\phi}\Big[\left(
3\sqrt{2}\cos(2k\hat{x})+4\cos(2\phi)\right)\sin(2\phi)\Big]\label{FP3} \nonumber
\end{eqnarray}
As evident in the jump operators $\mathbf{L}$, we see the desired behavior that a decay out of either the left- or right-going ports is associated with \emph{both} a left and right momentum kick. This feature results from the mirrors eliminating most of the ``which-way" information in the outgoing photon. However, we are missing a critical component: a way for the emitter to become re-excited by the photonic component still in the cavity---the Jaynes-Cummings mechanism. Without explicit quantization of the trapped mode between the mirrors, this behavior will not appear in the SLH description of this quantum network. 
Thus the resulting SLH-triple (\ref{FP1})-(\ref{FP3}), are not a valid description of the dynamics when the trapped optical mode is significantly populated \cite{Tabak2015, CookSchuClel18}. 

\subsection{Reduction of internal connections in SLH networks}\label{Reduction}
We now give some details regarding the steps on how to implement the reduction of the internal links in the input-output network shown in Fig \ref{FPSetup}. We first note that the more abstract depiction in Fig \ref{FPSetup}b) although having only two real inputs  and output (solid arrows), the beam splitters and emitter together have six inputs and outputs. Most of these are internally linked e.g.  output port \# 1 is routed to input port \# 6, output \# 6 goes to input \# 3 etc. As one eliminates each internal connection the number of input and output ports reduces by one each, e.g. after one reduction one is left with a 5 input : 5 output device. After each reduction one must take care to re-identify the enumeration of the ports. As an example we can track the enumeration of the reducing network as we eliminate (in order), the internal links $1_{out}\rightarrow 6_{in}\;:\; 6_{out}\rightarrow 3_{in}\;:\; 4_{out}\rightarrow 5_{in}\;:\; 5_{out}\rightarrow 2_{in}$, where we have used the labelling in Fig \ref{FPSetup}b). The actual enumeration of input to output ports must collapse as the size of the device shrinks and we give such a reducing enumeration for the example elimination we described in the previous sentence in Table \ref{Ports}.
\begin{table}
\begin{center}
\begin{tabular}{ccccccc|ccccc}
In &    &    &    &    &   &   & Out& & & &   \\
1  & 1 & 1 & 1 & 1 &   &   & \cancel{1} & & & &\\
2  & 2 & 2 & \cancel{2} &    &   &   & 2 & 1 & 1 & 1 & 1\\
3   &  \cancel{3} & & & & & & 3 & 2 & 2 & 2 & 2 \\
4 & 4 & 3 & 3 & 2& & & 4 & 3 & \cancel{3} & & \\
5 & 5 & \cancel{4} & & & & & 5 & 4 & 4 & \cancel{3} & \\
\cancel{6} & & & & & & & 6 &\cancel{5} & & &\\
\end{tabular}
\caption{Reduction of enumerating the input and output ports as one eliminates internal links in the network in Fig \ref{FPSetup}b). To derive this table we proceed as follow: the left most columns enumerate the original input(left) and output(right) ports. We first eliminate the internal link connecting $1_{out}\rightarrow 6_{in}$. We then relabel the nodes in the second to left most columns of the remaining ports which now range from $1:5$. We next eliminate the returning link, which in the new enumeration links $5_{out}\rightarrow 3_{in}$. We proceed to eliminate all four internal links to arrive at the reducing link identifications: $1_{out}\rightarrow 6_{in}\;: \; 5_{out}\rightarrow 3_{in}\;:\; 3_{out}\rightarrow 4_{in}\;:\; 3_{out}\rightarrow 2_{in}$. After these four reductions one is left with a 2 input, 2 output port device.}
\end{center}\label{Ports}
\end{table}

We now indicate how to perform the first reduction, obtaining the reduced network once the internal link $1_{out}\rightarrow 6_{in}$, is eliminated. We refer the reader to the SLH composition rules in section 5.2 of \cite{Combes2016}, and in particular Rule 4 or Eqns (61) of \cite{Combes2016}. This rules describes how to obtain the reduced  $G_{\rm red} = (\mathbf{S}_{\rm red}, \mathbf{L}_{\rm red}, \hat{H}_{\rm red})$ description from the original network $G= (\mathbf{S}, \mathbf{L}, \hat{H})$, when the internal link connecting output port $x$ to input port $y$, i.e. $x\,\rightarrow \,y$, is eliminated.  In our example, $G$ describes a $6\times 6$, input-output device while $G_{\rm red}$ describes a $5\times 5$ device, and we have
\begin{eqnarray}
{\bf S}_{\rm red} &=& {\bf S}_{\bar{x}\bar{y}}+{\bf S}_{\bar{x}y}(I-S_{xy})^{-1}{\bf S}_{x\bar{y}}\;\;,\label{S61a}\\
{\bf L}_{\rm red}&=& {\bf L}_{\bar{x}}+{\bf S}_{\bar{x}y}(I-S_{xy})^{-1}L_x\;\;,\label{S61b}\\
\hat{H}_{\rm red}&=&\hat{H}+\frac{1}{2i}\left({\bf L}^\dagger{\bf S}_{:,y}(1-S_{xy})^{-1}L_x-{\it c.c.}\right)\;\;,\label{S61c}
\end{eqnarray}
where ${\bf S}_{\bar{x}\bar{y}}$ is the scattering matrix ${\bf S}$, omitting the $x^{th}$ row and $y^{th}$ column. ${\bf S}_{\bar{x}y}$ is the $y^{th}$ column of $\bf S$ with the $x^{th}$ row deleted,   ${\bf S}_{x\bar{y}}$ is the $x^{th}$ row of $\bf S$ with the $y^{th}$ column deleted, and $S_{xy}$ is the $(x,y)$ element of ${\bf S}$. Also, ${\bf L}_{\bar x}$ is $\bf L$ without the $x^{th}$ row, and $L_x$ is the $x^{th}$ row of $\bf L$.

We begin with the full description of $G = (\mathbf{S}, \mathbf{L}, \hat{H})$ for the $6 \times 6$ network:
\begin{equation}
{\bf S}=\left(
\begin{array}{cccccc}
 i \cos (\beta ) & \sin (\beta ) & 0 & 0 & 0 & 0 \\
 \sin (\beta ) & i \cos (\beta ) & 0 & 0 & 0 & 0 \\
 0 & 0 & i \cos (\alpha ) & \sin (\alpha ) & 0 & 0 \\
 0 & 0 & \sin (\alpha ) & i \cos (\alpha ) & 0 & 0 \\
 0 & 0 & 0 & 0 & e^{i \left(\phi _1+\phi _2\right)} & 0 \\
 0 & 0 & 0 & 0 & 0 & e^{i \left(\phi _1+\phi _2\right)} \\
\end{array}
\right)\;,
\label{S6}
\end{equation}
\begin{equation}
{\bf L}=\sqrt{\frac{\Gamma}{2}} \hat{\sigma}_-\,\left(
\begin{array}{c}
 0 \\
 0 \\
 0 \\
 0 \\
 e^{i k \hat{x}+i \phi_2} \\
 e^{i \phi_1-i k \hat{x}}  \\
\end{array}
\right)\;,\label{L6}
\end{equation}
\begin{equation}
\hat{H} = \hat{H}_{\rm sys}= \omega_m \hat{v}\dg \hat{v} \label{H6}\;.
\end{equation}

We now eliminate the link $1 \,\rightarrow 6$, to obtain
\begin{equation}
{\bf S}_{\rm red}=
\left(
\begin{array}{ccccc}
 \sin (\beta ) & i \cos (\beta ) & 0 & 0 & 0 \\
 0 & 0 & i \cos (\alpha ) & \sin (\alpha ) & 0 \\
 0 & 0 & \sin (\alpha ) & i \cos (\alpha ) & 0 \\
 0 & 0 & 0 & 0 & e^{i \left(\phi _1+\phi _2\right)} \\
 i e^{i \left(\phi _1+\phi _2\right)} \cos (\beta ) & e^{i \left(\phi _1+\phi _2\right)} \sin (\beta ) & 0 & 0 & 0 \\
\end{array}
\right)\label{S5} ,
\end{equation}
while 
\begin{equation}
{\bf L}_{\rm red}=\sqrt{\frac{\Gamma}{2}} \hat{\sigma}_-\,
\left(
\begin{array}{c}
 0 \\
 0 \\
 0 \\
 e^{i \phi_2 + i k \hat{x}}  \\
 e^{i \phi _{1}-i k \hat{x}}  \\
\end{array}
\right) , 
\label{L5}
\end{equation}
and $\hat{H}_{\rm red}=\hat{H}_{\rm sys}$, since $L_1=0$, and $S_{16}=0$.
To perform the next reduction we must refer back to the Fig \ref{FPSetup}b). In the original enumeration of the $6\times 6$ device we next eliminate the internal link $6\rightarrow 3$, but in the new labeling of the $5\times 5$ reduced device we refer to Table \ref{Ports}, and using the second to left most column of the Out and In ports we next eliminate the internal $5\rightarrow 3$ link. We follow the above procedure to finally arrive at a $2 \times 2$ SLH network.

\end{document}